\documentclass[showpacs,prl,onecolumn,aps,superscriptaddress,preprintnumbers]{revtex4}
\usepackage{amsmath,amssymb}
\usepackage{epsfig}
\usepackage{graphicx}
\usepackage{amsmath}
\usepackage{amsfonts}
\usepackage{epstopdf}
\def\slashchar#1{\setbox0=\hbox{$#1$}     		
   \dimen0=\wd0                                 	
   \setbox1=\hbox{/} \dimen1=\wd1               	
   \ifdim\dimen0>\dimen1                        	
      \rlap{\hbox to \dimen0{\hfil/\hfil}}      	
      #1                                        	
   \else                                        	
      \rlap{\hbox to \dimen1{\hfil$#1$\hfil}}   	
      /                                         	
   \fi}

\newcommand{\eval}[1]{\left \langle #1 \right \rangle}

\renewcommand{\vec}{\boldsymbol}
\newcommand{\be}{\begin{equation}}
\newcommand{\ee}{\end{equation}}
\newcommand{\bea}{\begin{eqnarray}}
\newcommand{\eea}{\end{eqnarray}}
\newcommand{\ba}{\begin{array}}
\newcommand{\ea}{\end{array}}

\def\eq#1{{Eq.~(\ref{#1})}}
\def\fig#1{{Fig.~\ref{#1}}}

\newcommand{\as}{\alpha_S}
\newcommand{\nn}{\nonumber}

\newcommand{\Lb}{\left(}
\newcommand{\Rb}{\right)}
\newcommand{\h}{\frac{1}{2}}

\begin{document}

\title{Color confinement and screening in the $\mathbf{\theta}$ -vacuum}

\author{Dmitri E. Kharzeev}
\email{Dmitri.Kharzeev@stonybrook.edu}
\affiliation{Department of Physics and Astronomy, Stony Brook University, New York 11794-3800, USA}
\affiliation{Department of Physics and RIKEN-BNL Research Center, \\
Brookhaven National Laboratory, Upton, New York 11973-5000, USA}
\author{Eugene M. Levin}
\email{leving@post.tau.ac.il, eugeny.levin@usm.cl}
\affiliation{Department of Particle Physics, School of Physics and Astronomy,
Tel Aviv University, Tel Aviv, 69978, Israel}
\affiliation{Departamento de F\'\i sica,
Universidad T$\acute{e}$cnica Federico Santa Mar\'\i a   and
Centro Cient\'\i fico-Tecnol$\acute{o}$gico de Valpara\'\i so,
Casilla 110-V,  Valparaiso, Chile}

\date{\today}


\begin{abstract}
QCD perturbation theory ignores the compact nature of $SU(3)$ gauge group that gives rise to the periodic $\theta$-vacuum of the theory. We propose to modify the gluon propagator to reconcile perturbation theory with the anomalous Ward identities for the topological current in the $\theta$-vacuum. As a result, the gluon couples to the Veneziano ghost describing the tunneling transitions between different Chern-Simons sectors of the vacuum; we call the emerging gluon dressed by ghost loops a ``glost".  We evaluate the glost propagator and find that it has the form $G(p) = (p^2 + \chi_{top}/p^2)^{-1}$ where $\chi_{top}$ is the Yang-Mills topological susceptibility related to the $\eta'$ mass by Witten-Veneziano relation; this propagator describes confinement of gluons at distances $\sim \chi_{top}^{-1/4} \simeq 1$ fm. The same functional form of the propagator was originally proposed by Gribov as a solution to the gauge copies problem that plagues perturbation theory.  The resulting running coupling coincides with the perturbative one at $p^2 \gg \sqrt{\chi_{top}}$, but in the infrared region either freezes (in pure Yang-Mills theory) or vanishes (in full QCD with light quarks), in accord with experimental evidence. Our scenario makes explicit the connection between confinement and topology of the QCD vacuum; we discuss the implications for spin physics, high energy scattering, and the physics of quark-gluon plasma.  
\end{abstract}
\maketitle


QCD possesses a compact gauge group that allows for topologically non-trivial gauge field configurations. These configurations realize homotopy maps from the gauge group to the space-time manifold. For example, the homotopy map from the ${\rm SU(2)}$ subgroup of the gauge group to the Euclidean space-time sphere ${\rm S^3}$ describes the instanton solution \cite{Belavin:1975fg}. However the compactness of the gauge group is ignored in perturbation theory, and this may be at the origin of problems marring the perturbative approach. 
\vskip0.3cm
In QCD, one of these problems is the existence of Gribov copies \cite{GRCO} -- multiple solutions of the gauge-fixing condition that make the perturbative approach ambiguous. In Coulomb gauge, the emergence of Gribov copies can be traced back to the existence of energy-degenerate vacua with different Chern-Simons numbers \cite{Jackiw:1977ng}. A natural question arises -- is it possible to formulate QCD perturbation theory in a way that is consistent with the topological structure of the theory? In this Letter we argue that the answer to this question is positive. We find that the resulting gluon propagator naturally describes confinement, i.e. non-propagation of color degrees of freedom, and the running coupling displays the screening of color charge at large distances. 

\vskip0.3cm
 In Minkowski space-time, instanton solutions represent the tunneling events connecting the degenerate vacuum states with different Chern-Simons numbers 
\be\label{CS_num}
X (t) = \int d^3 x\ K_0 (x, t),
\ee
where $K_0$ is the temporal component of topological current 
\be\label{CS_cur}
K_\mu = \frac{g^2}{16 \pi^2} \epsilon_{\mu\nu\rho\sigma} A^{\nu,a} \left(\partial^\rho A^{\sigma,a} + \frac{1}{3}g C^{abc} A^{\rho}_b A^{\sigma}_c \right) ;
\ee 
the first term in $K_0$ is the density of Abelian ``magnetic helicity" while the second term is its non-Abelian generalization.

The chiral anomaly in QCD leads to non-conservation of the axial current
\be
\partial_\mu J_A^\mu = 2 N_f Q(x) + \sum_f (2 i m_f) \bar{q}_f \gamma_5 q_f,
\ee
where $m_f$ are the masses of quarks, $N_f$ is the number of flavors, and 
\be
Q(x) = \frac{g^2}{32 \pi^2} F_{\mu\nu}(x) \tilde{F}^{\mu\nu}(x) 
\ee
is the density of topological charge normalized by $\int d^4x\ Q(x) = \nu$; for finite action field configurations $\nu$ is an integer. 
The density of topological charge can be represented as a divergence $Q(x) = \partial^\mu K_\mu$ of the gauge-dependent current (\ref{CS_cur}).
\vskip0.3cm

Veneziano \cite{VEN} has demonstrated that the periodic $\theta$-vacuum stucture in QCD can be captured by introducing a massless ``ghost" in the correlation function of the gauge-dependent topological current (\ref{CS_cur}):
\be \label{VEN}
K_{\mu \nu}\Lb q \Rb\,\equiv \,i \,\int d^4 x\ e^{ i q x}\,\eval{K_\mu\Lb x\Rb K_\nu \Lb 0\Rb}\,\,\xrightarrow{ q^2 \,\ll \,\mu^2}\,\,- \frac{\mu^4}{q^2}\ g_{\mu \nu},
\ee
where $\mu^4 \equiv \chi_{top}$ is the topological susceptibility of pure Yang-Mills theory. Note that the r.h.s. of \eq{VEN} has the ``wrong" sign, i.e. the ghost does not describe a propagating degree of freedom. This means that the ghost cannot be produced in a physical process; however the couplings of the ghost (that describe the effect of topological fluctuations) certainly can affect physical amplitudes. A similar ``dipole" ghost had been earlier introduced by Kogut and Susskind \cite{Kogut:1974kt} in the analysis of axial anomaly in the Schwinger model. This procedure has been demonstrated to solve the $U_A(1)$ problem in QCD \cite{Weinberg:1975ui,Witten:1979vv}.
\vskip0.3cm

The physical meaning of \eq{VEN} becomes apparent if one compares it to the correlation function of the electron's coordinate $x(t)$ in a crystal \cite{Dyakonov_Eides}:
\be\label{cryst}
i \int dt \ e^{i \omega t}\ \langle T\{ x(t) x(0) \}\rangle \  \xrightarrow{\omega \to 0} - \frac{1}{\omega^2\ m^*} = - \frac{1}{\omega^2}\  \frac{\partial^2 E(k)}{\partial k^2} \biggr\rvert_{k = 0} ,
\ee
where $E = k^2/ 2 m^*$ is the energy of an electron with an effective mass $m^*$ and quasi-momentum $k$ in a crystal. The emergence of the pole in \eq{cryst} signals the possibility of electron's propagation in the periodic potential of the crystal due to tunneling. Note that the pole emerges not just from a single tunneling event (corresponding to the instanton in QCD), but sums up the effect of many tunnelings throughout the crystalline lattice. 
\vskip0.3cm
The analogy between \eq{VEN} and \eq{cryst} can be made even more apparent if we choose the frame with $q^\mu = (\omega, 0)$ and use the (gauge-invariant) analog of coordinate given by \eq{CS_num}. 
The expression \eq{VEN} then takes the form completely analogous to \eq{cryst}:
\be
i \int dt\ e^{ i \omega t}\,\langle T\{X(t) X(0)\} \rangle \,\,\xrightarrow{\omega \to 0} - \frac{\mu^4}{\omega^2}\ V = - \frac{1}{\omega^2}\ \frac{\partial^2 E(\theta)}{\partial \theta^2} \biggr\rvert_{\theta = 0} ,
\ee
where $V$ is the volume of the system, and $E(\theta) = \epsilon(\theta) V$ is the energy of the Yang-Mills vacuum. 
The energy density of the vacuum $\epsilon(\theta)$ is a periodic function of the $\theta$ angle that is analogous to the quasi-momentum $k$ in \eq{cryst}. At small $\theta$, we can expand $\epsilon(\theta)$ and write 
\be
\epsilon(\theta) = \mu^4\ \frac{\theta^2}{2} ,
\ee 
which exhibits the physical meaning of $\mu^4$ as of the topological susceptibility $\chi_{top} = \mu^4$ of the Yang-Mills theory; note that a term linear in $\theta$ is forbidden by P and CP invariances of QCD.
\vskip0.3cm

It is well known that topological susceptibility vanishes, order by order, in perturbation theory. Perturbative description thus corresponds to $\mu \to 0$ in \eq{VEN}, or to the limit of the infinitely heavy electron, $m^* \to \infty$ in \eq{cryst}.  Infinitely heavy electrons do not respond to electromagnetic fields, as the corresponding coupling of electromagnetic current $j = e\ q/m^*$ to the gauge field $j A \sim 1/m^*$ vanishes in the limit $m^* \to \infty$. In this case the dynamics of photons is not sensitive to the periodic structure of the crystal, and one can build the usual perturbation theory of photons. However when $m^*$ is finite, and is of the order of the frequency of the external gauge field, photons can be absorbed and re-emitted by the electrons, and these processes severely affect the photon propagator. Also, at finite $m^*$, the static Coulomb field can be screened by the electrons at large distances.

\vskip0.3cm
As we will discuss below, the situation in QCD is very similar -- as $\mu \to 0$, the periodic structure of the $\theta$ vacuum becomes irrelevant. However, in the physical world $\mu \sim \Lambda_{\rm QCD} \sim 200$ MeV, so the ghost (describing the tunneling in the periodic $\theta$-vacuum) strongly affects propagation of gluons with frequencies $\omega \sim \mu$. At large distances, the ghost also gives rise to the screening of color charge, leading to the freezing of the effective coupling in the infrared ${\rm IR}$ limit for pure gauge theory, or to the  vanishing of the coupling in the ${\rm IR}$ in QCD with light quarks. Because $\mu \sim \Lambda_{\rm QCD} \sim 1/R_{conf}$ is on the order of inverse confinement radius $R_{conf}$, these phenomena describe confinement of gluons at distances $R_{conf} \simeq 1$ fm.

\begin{figure}
\begin{center}
\vspace{-6cm}
\includegraphics[width=14cm]{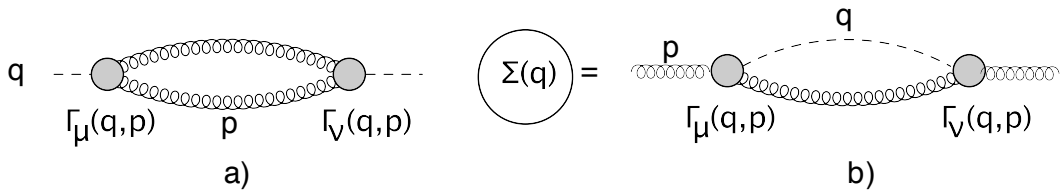}
\end{center}
\vspace{-7cm}
\caption{\fig{kk}-a : \eq{VEN} in momentum representation; helix lines represent gluons and the dashed line depicts the ghost. \fig{kk}-b : the gluon dressed by the interactions with the ghost: a ``glost", see \eq{SIGMA} for the corresponding self-energy expression.}
\label{kk}
\end{figure}

\vskip0.3cm
The key observation of our paper is that \eq{VEN} and \eq{CS_cur} define an effective ghost-gluon-gluon vertex  $\Gamma_\mu\Lb q,p\Rb$. Using this vertex, we can re-write the correlator  \eq{VEN} at small $q^2$ as follows (see \fig{kk}-a): 
\be \label{KMOM}
K_{\mu \nu}\Lb q \Rb\,\,=\,\,\frac{1}{( 2 \pi)^4 i}\int d^4 p ~\Gamma_\mu\Lb q,p\Rb\frac{1}{ p^2 (q -p )^2}\, \Gamma_\nu\Lb q,p\Rb\,=\,\,- \frac{\mu^4}{q^2}\,g_{\mu \nu}
\ee

From \eq{KMOM} we find that
\be \label{GG}
 \Gamma_\mu\Lb q,p \Rb\,\Gamma_\nu\Lb q,p\Rb\,\,\propto\, - \frac{\mu^4}{p^2}\,g_{\mu \nu}\ ,\,\mbox{for}\,\,q \,\leq p .
 \ee 
The vertices  $\Gamma_\mu\Lb q,p\Rb$ describe the excitation of the ghost by gluons, and affect the gluon propagation at small virtualities. 
 
\begin{figure}
\begin{center}
\vspace{-5cm}
\includegraphics[width=12cm]{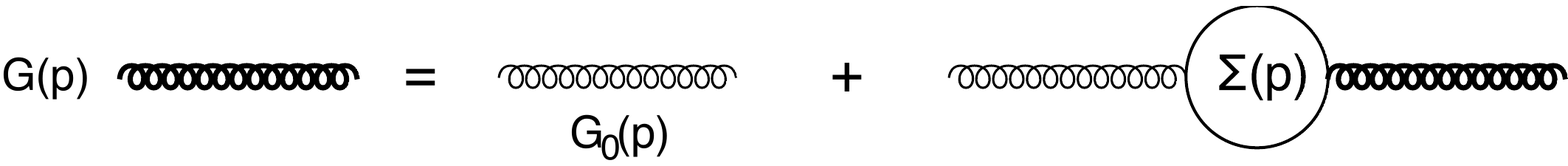}
\end{center}
\vspace{-6cm}
\caption{The graphic form of the equation for the gluon propagator, \eq{EQ}. The full propagator is denoted by the bold helix line; thin helix stands for the perturbative gluon propagator $G_0(p)$, and the blob for self-energy $\Sigma(p)$ given by \eq{SIGMA} and \eq{SIGMA1}.}
\label{prop}
\end{figure}

Indeed, the gluon  propagator is now the solution of the equation shown in \fig{prop} with $\Sigma\Lb p\Rb$
given by 

\be \label{SIGMA}
\Sigma_{\mu\nu}\Lb p\Rb~=~\frac{1}{( 2 \pi)^4 i}\int d^4 q ~\Gamma_\mu\Lb q,p\Rb\frac{1}{ q^2 (q -p )^2}\, \Gamma_\nu\Lb q,p\Rb
\ee
where $1/q^2$ is the propagator of the ghost. In evaluating the integral of \eq{SIGMA} we assume that $q \ll\,p$ since 
$\Gamma_\mu$'s describe the non-perturbative effects concentrated at small momenta (long distances). Therefore,
\be \label{SIGMA1}
\Sigma_{\mu\nu}\Lb p\Rb~\simeq~\frac{4\pi }{( 2 \pi)^4 }\int^p_0 d q^2 q^2 ~\Gamma_\mu\Lb q,p\Rb\frac{1}{ q^2 p^2}\, \Gamma_\nu\Lb q,p\Rb\,\,= \,\,-\,g_{\mu \nu}\,\frac{\mu^4}{p^4}\int ^{p^2} d q^2  \,\,=\,\,- g_{\mu \nu} \frac{\mu^4}{p^2} ,
\ee
where we used \eq{GG}; 
note that in \eq{SIGMA1} we made rotation to  the pseudo-Euclidean space. We can now write down the Schwinger-Dyson equation for the gluon propagator \footnote{This is the expression in Feynman gauge; since the self-energy $\Sigma_{\mu\nu}$ does not change the polarization of the gluon, the Dyson-Schwinger series can be re-summed in the usual way.} $G_{\mu \nu}\Lb p\Rb\,\,=\,\,g_{\mu \nu}\,G\Lb p\Rb$ in terms of $\Sigma_{\mu\nu}\Lb p\Rb = g_{\mu \nu}\, \Sigma\Lb p\Rb$, see Fig. \ref{prop} :
\be \label{EQ}
G\Lb p \Rb\,\,=\,\,\frac{1}{p^2} \, + \,\frac{1}{p^2} \Sigma\Lb p\Rb \,G\Lb p \Rb
\ee
with the solution
\be \label{SOL1}
G\Lb p \Rb\,\,=\,\,\frac{1}{p^2 - \Sigma\Lb p\Rb}\,\,=\,\,\frac{1}{ p^2 + \frac{\mu^4}{p^2}}
\ee
\vskip0.3cm

The propagator (\ref{SOL1}) has remarkable properties. First, 
   $G\Lb p \Rb$ has no infrared singularities and no gluon pole in the physical region. Indeed, this propagator has only complex poles at $p^2 \,=\,\pm i \mu^2$.  As a result, gluons cannot be observed as particles in detectors -- in other words, they are confined. 
   Second, the propagator of the type of \eq{SOL1}  was proposed by Gribov \cite{GRCO} as a solution to the problem of gauge copies -- multiple solutions to the gauge fixing condition, see \cite{GRCREV,Dokshitzer:2004ie} for reviews. Hence we can state that introducing the coupling to the ghost (and thus taking account of the periodicity of the $\theta$-vacuum) solves the problem of Gribov copies and leads to the confinement of gluons. The dimensionful Gribov parameter acquires a well-defined meaning of topological susceptibility $\chi_{top} = \mu^4$ related to the $\eta'$ mass by the Witten-Veneziano relation \cite{VEN,Witten:1979vv}; since $\mu \simeq \Lambda_{\rm QCD}$, confinement emerges at distances of about 1 fm. Note that close to the deconfinement transition, the topological susceptibility vanishes reflecting the restoration of $U_A(1)$ symmetry \cite{Bonati:2013tt,Kharzeev:1998kz}, see \cite{Kharzeev:2015kna} for a review. Since at $\mu \to 0$ the gluon propagator becomes perturbative, the restoration of $U_A(1)$ symmetry and deconfinement should occur at the same temperature as suggested by the lattice data \cite{Bonati:2013tt}; however close to $T_c$ the non-perturbative interactions induced by $\mu \neq 0$ are important.
\vskip0.3cm   
   
In our approach, the propagator \eq{SOL1} results from the admixture of the ghost to the perturbative gluon (see \fig{kk}-b), with an amplitude defined by the topological susceptibility $\mu^4$.  We thus propose the following name for the particle with propagator given by \eq{SOL1} that represents a coherent mixture of a {\bf gl}uon and a gh{\bf ost} -- a ``{\bf glost}" \footnote{``Urban dictionary" gives the following definition of this word: ``Glost is a multi-versatile word used to explain something positive or desirable, generally something that occurs between two close friends"; a more traditional meaning is the ``glaze fused to ceramic ware".}.  Unlike the ghost, the ``glost" can be produced in a physical process, but unlike the perturbative gluon, it is confined and can propagate only at short distances $\sim \mu^{-1} \sim 1$ fm.
\vskip0.3cm

\begin{figure}
\begin{center}
\vspace{1cm}
\includegraphics[width=14cm]{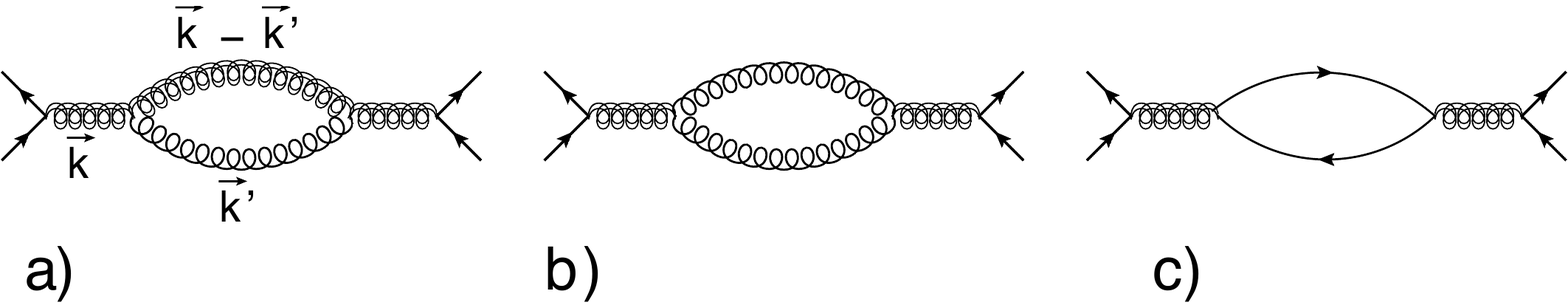}
\end{center}
\caption{The first order corrections to the Coulomb energy. Helix lines denote the transversely polarized gluon, the double helix lines show the longitudinally polarized gluon, and the solid arrow lines depict the quarks.}
\label{run}
\end{figure}

Let us now re-consider the asymptotic freedom of QCD \cite{Gross:1973id,Politzer:1973fx} using the ``glost" propagator \eq{SOL1}. 
 In our derivation we will compute the interaction energy of two heavy quarks in the Coulomb gauge \cite{KHR,GRIB} that is free from the Faddeev-Popov ghosts, so we can avoid dealing with two different types of ghosts.  The dominant contribution responsible for the asymptotic freedom stems from the diagram of \fig{run}-a. The contribution of this digram in perturbative QCD takes the form
 \be \label{pQCD1}
 \Pi^{(a)}\,\,=\,\,3 g^2 C^2_2 \int \frac{d^4 k'}{(2 \pi)^4 i}\frac{1}{( \vec{k} - \vec{k}')^2\,(k'^2_0 - \vec{k}'^2)}\Big( 1 - \frac{\Lb \vec{k} \cdot \vec{k}'\Rb^2}{\vec{k}^2 \,\vec{k}'^2}\Big) ;
 \ee
$\Pi$ is related to $\Sigma$ by $\Sigma = k^2 \Pi$.
 
 For the ``glost" propagator (\ref{SOL1}), \eq{pQCD1} changes and takes a different form: 
 \bea \label{QCD2}
 \Pi^{(a)}\,\,&=&\,\,3 g^2 C^2_2 \int \frac{d^4 k'}{(2 \pi)^4 i}\frac{( \vec{k} - \vec{k}')^2}{\Lb( \vec{k} - \vec{k}')^2\Rb^2 + \mu^4} \,\frac{(k'^2_0 - \vec{k}'^2)}{\Lb(k'^2_0 - \vec{k}'^2) \Rb^2 + \mu^4}
 \Big( 1 - \frac{\Lb \vec{k} \cdot \vec{k}'\Rb^2}{\vec{k}^2 \,\vec{k}'^2}\Big)\nn\\
 &=&\,\,3 g^2 C^2_2 \int \frac{d^4 k'}{(2 \pi)^4 i} \mbox{Re}\Bigg\{ \frac{1}{( \vec{k} - \vec{k}')^2 + i \mu^2}\Bigg\}\,\mbox{Re}\Bigg\{ \frac{1}{k'^2_0 -  \vec{k}'^2 + i \mu^2}\Bigg\}\frac{1}{\vec{k}^2 \vec{k}'^2}  
 \Lb \vec{k}^2 \vec{k}'^2\,-\,\Lb \vec{k} \cdot \vec{k}'\Rb^2\Rb 
  \eea

In \eq{QCD2} we have four terms obtained by choosing different signs of $i \mu^2 $. To illustrate the procedure of calculation, let us evaluate one of them:
\be \label{QCD3}
 \Pi^{(a)}_1\,\,=3 g^2 C^2_2 \int \frac{d^4 k'}{(2 \pi)^4 i} \frac{ \vec{k}^2 \vec{k}'^2\,-\,\Lb \vec{k} \cdot \vec{k}'\Rb^2}{\Lb ( \vec{k} - \vec{k}')^2 + i \mu^2\Rb \Lb k'^2_0 -  \vec{k}'^2 + i \mu^2\Rb\vec{k}^2 \vec{k}'^2}   .
 \ee
 Introducing Feynman parameters $\alpha_1 + \alpha_2 + \alpha_3=1$ we obtain ($\vec{P} = \vec{k}' - \alpha_1 \vec{k}$ and $\as = g^2/4\pi$):
 \bea \label{QCD4}
  \Sigma^{(a)}_1\,\,&=&\,\, 3\,g^2 C^2_2 \frac{1}{\vec{k}^2}\,\int^1_0 d \alpha_1 \int^{1 - \alpha_1} _0 d \alpha_2  \int \frac{d^4 k'}{(2 \pi)^4 i} \frac{ \vec{k}^2 \vec{P}^2 - \Lb  \vec{k}\cdot \vec{P}\Rb^2}{\Lb \alpha_2 k^2_0 - \vec{P}^2 - k^2\alpha_1 (1 - \alpha_1) + i \mu^2(\alpha_1 + \alpha_2)\Rb^3}\nn\\
&  \xrightarrow{\mbox{integrating over $k'_0$}}&\,\, \,g^2 C^2_2 \int^1_0 \frac{d \alpha_2}{\sqrt{\alpha_2}} \int^{1 - \alpha_2} _0 d \alpha_1  \int \frac{d^3 \vec{P}}{(2 \pi)^3} \frac{3}{16}\frac{\vec{P}^2}{\Lb\vec{P}^2 + k^2*\alpha_1 (1 - \alpha_1) +i \mu^2 ( \alpha_1+ \alpha_2)\Rb^{5/2}}\nn\\
 &  \xrightarrow{\mbox{integrating over $\vec{P}$}}&  \frac{ 3 \as}{ 8 \pi} C^2_2 \int^1_0 \frac{d \alpha_2}{\sqrt{\alpha_2}} \int^{1 - \alpha_2} _0 d \alpha_1\Bigg\{ - \frac{4}{3}  + \ln\Lb \frac{2 L}{\sqrt{k^2 \alpha_1 (1 - \alpha_1) + i \mu^2 \Lb \alpha_1 + \alpha_2\Rb}}\Rb\Bigg\} ,
 \eea
 where $L$ is an ultraviolet cutoff in the integration over momentum. The integrals over $\alpha_1$ and $\alpha_2$ can be taken analytically. The main features of  \eq{QCD4} are the following: it has a logarithmic divergence at large $k$ and is finite at $k=0$.

 Summing all terms we find that we need to replace $ \ln\Lb L^2/k^2\Rb$ of perturbative QCD in the diagram of Fig. \ref{run}-a by the following function: 
 \bea \label{QCD5}
&&  \ln\Lb L^2/k^2\Rb \,\longrightarrow\, \ln\Lb L,k,\mu\Rb\,\equiv\,\frac{3}{8} \int^1_0 \frac{d \alpha_1}{\sqrt{\alpha_1}} \int^{1 - \alpha_1} _0 d \alpha_2\nn\\
&&\Bigg\{ - \frac{16}{3}  + \ln\Lb \frac{4 L^2}{\sqrt{k^4 \alpha^2_2 (1 - \alpha_2)^2 +  \mu^4 \Lb \alpha_1 + \alpha_2\Rb^2}}\Rb\,+\,\ln\Lb \frac{4 L^2}{\sqrt{k^4 \alpha^2_2 (1 - \alpha_2)^2 + \mu^4 \Lb \alpha_1 -\alpha_2\Rb^2}}\Rb\Bigg\}
  \eea
  
  It turns out that the same substitution has to be done in the expression for \fig{run}-b that in perturbative approach gives a positive contribution to the $\beta$-function. However the contribution of the quark loop (\fig{run}-c) remains the same as in perturbative QCD. The sign of this contribution to the $\beta$-function is also positive, and it leads to Landau pole and a ``Moscow zero" \cite{LAKH} . As a result the QCD coupling in our approach tends to zero in the infrared region of $k \to 0$. 
  \vskip0.3cm
  
\begin{figure}
  \vspace{-2cm}
\begin{tabular}{ c c c}
\includegraphics[width=7cm]{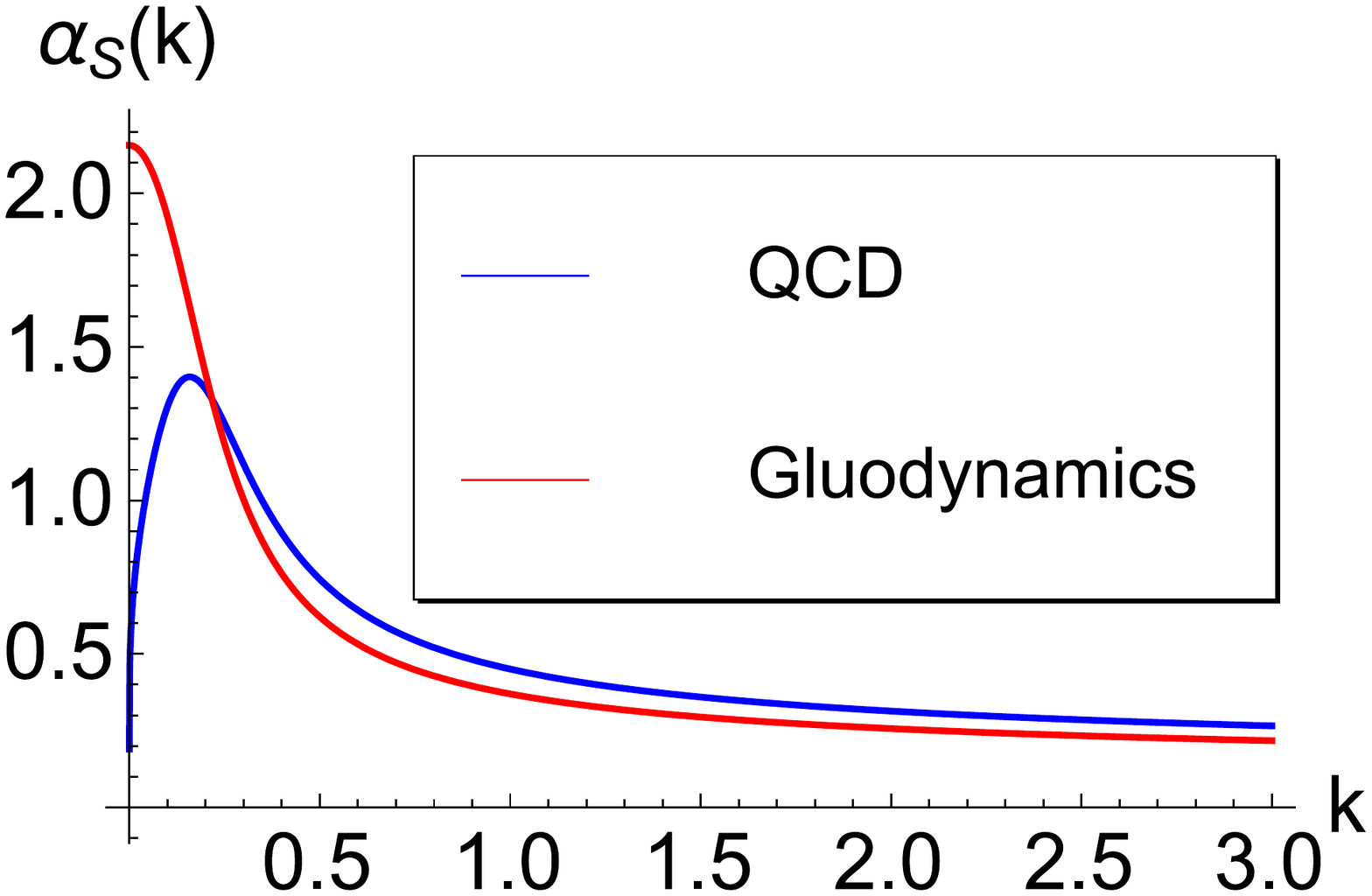} & ~~~~~~~~~&\includegraphics[width=7cm]{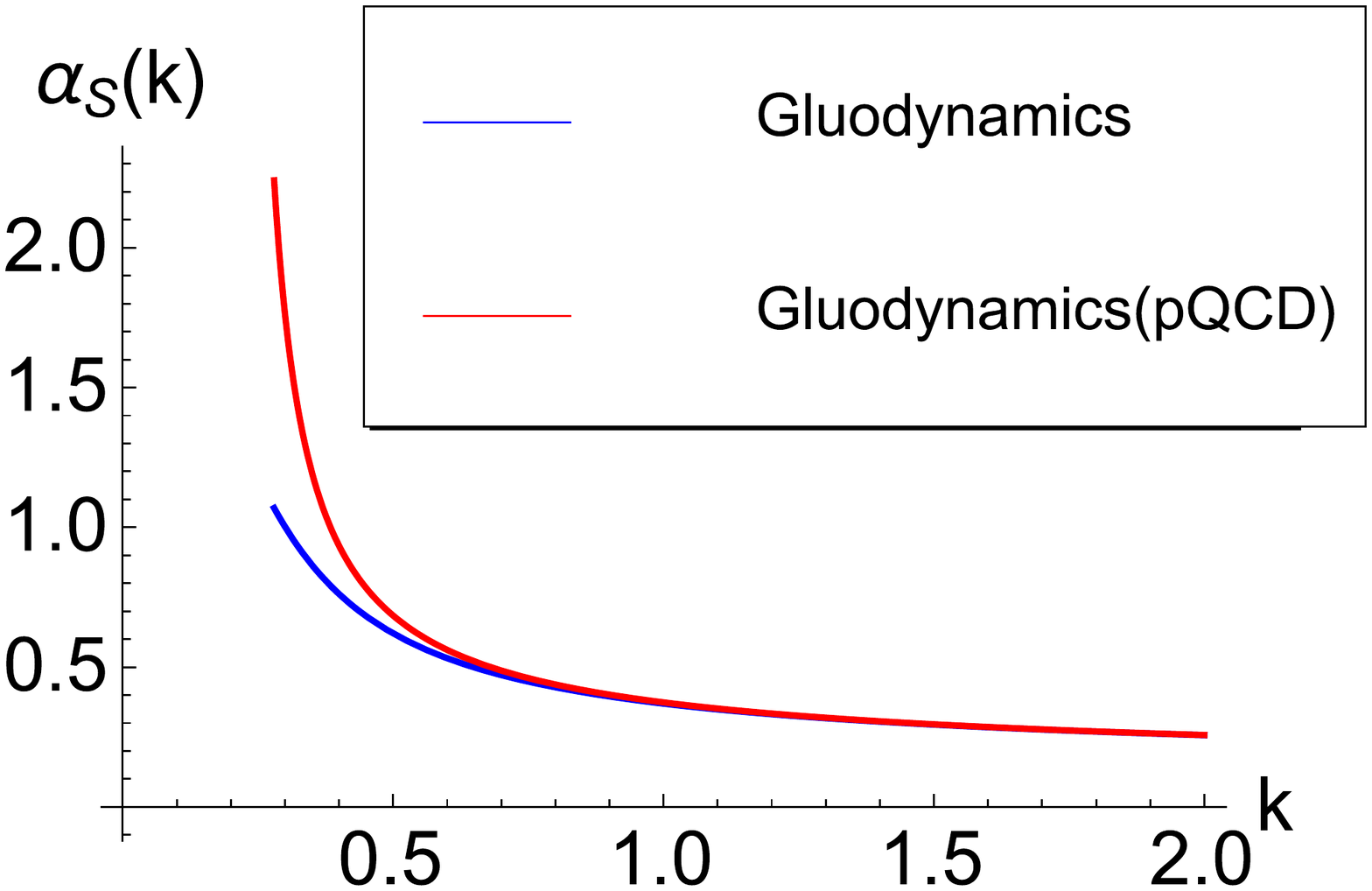}\\
\vspace{-2cm}
\end{tabular}
\caption{The  running constant $\as$ as a function of momentum $k$. \fig{as}-a shows our result for $\as$ in QCD with light quarks (blue line that goes to zero at $k\to0$) and gluodynamics (red line). \fig{as}-b shows the comparison of momentum dependence of $\as$ in gluodynamics in our approach (blue line) and perturbative QCD (red line that is above the blue one at small $k$). The renormalization point is chosen at the mass of Z-boson, $k = M_Z$.}
\label{as}
\end{figure}

  If we choose the renormalization point $k = \mu$, the running coupling takes the form 
  \be \label{QCD6}
  \as(k^2)\,\,=\,\,\frac{\as\Lb \mu\Rb}{1 + \as\Lb \mu\Rb \left[  \frac{11 N_c}{12 \pi} \,\Lb  \ln\Lb L,k,\mu\Rb - \ln\Lb
  L,\mu,\mu\Rb\Rb \,\,-\,\,\frac{2 N_f}{12 \pi}  \ln\Lb k^2/\mu^2\Rb\right]}
  \ee
  In \fig{as} we plot the coupling $\as$ as a function of $k$ for two cases: QCD and pure gluodynamics in our approach (\fig{as}-a), and the comparison of $\as$ in gluodynamics with perturbative QCD calculations in the leading order (\fig{as}-b). In both cases we choose the renormalization mass to be equal to the mass of the Z-boson; we use the value of $\mu = 0.18$ GeV from the original paper \cite{VEN} where it has been determined from the mass of $\eta'$-meson, and $N_c = N_f = 3$.

\vskip0.3cm
  
      One can see that replacing the gluon propagator by the propagator of the glost in gluodynamics removes the Landau pole and leads to the finite value of $\as$ at $k=0$.
  On the other hand, with the inclusion of quarks, the QCD coupling vanishes at $ k=0$.   At short distances, the running coupling is dominated by the perturbative contribution and so is not modified. The ghost affects the running coupling in a way that is quite different from the effect of a single instanton, which has been shown to increase the effective coupling at distances on the order of the instanton size \cite{Callan:1977gz,Randall:1998ra}. This may not be surprising as the ghost describes the effect of many instanton transitions throughout the $\theta$-vacuum. The screening effect of the ghost admixture is clearly a consequence of the fact that it is a spin-zero pseudoscalar ``particle".
 \vskip0.3cm
  
     It has been argued by Dokshitzer \cite{DOK} that the experimental data indicate that in the IR region the QCD coupling remains effectively small:  
     \be \label{QCD7}
    \alpha_0\,\,=\,\,\frac{1}{\mu_I}\,\int^{\mu_I} d k \,\as\Lb k \Rb\,\,\approx\,\,0.5 \,\,\,\,\,\mbox{for}\,\,\,\mu_I\,\,=\,\,2\,{\rm GeV} .
    \ee
 In  our approach we get $\alpha_0 \,=\,0.59$ for renormalization point $k = M_Z$, in reasonable agreement with \eq{QCD7}. 
\vskip0.3cm
It is of fundamental interest to establish the microscopic dynamics responsible for the long-range correlations of topological current captured by the ghost. A recent study within the ``deformed QCD" attributes these correlations to the topological order in the vacuum \cite{Zhitnitsky:2013hs}. Our result suggests a link between the confinement and the long-range topological correlations in the QCD vacuum, and provides a practical way of computing power-suppressed corrections to QCD amplitudes, in particular the ones that are forbidden in the perturbative approach. 

\vskip0.3cm
To summarize, we propose to modify the gluon propagator in perturbative QCD by taking account of the periodic structure of the QCD $\theta$-vacuum. Our prescription for the gluon propagator leads to the coupling of the gluons to the ghost saturating the anomalous Ward identity for topological current. The resulting ``glost" propagator appears to have the functional form originally proposed by Gribov, in which the role of dimensionful parameter is played by the topological susceptibility $\chi_{top} \equiv \mu^4$. Our approach thus removes the Gribov copies that usually plague perturbation theory, and describes confinement of gluons at distances $\sim \mu^{-1} \simeq 1$ fm. We also find that the running coupling in the {\rm IR} freezes in pure gauge theory, or tends to zero in QCD with light quarks. Because the topological susceptibility vanishes above the deconfinement transition, the ``glosts" become usual perturbative gluons in the deconfined phase at high temperatures. The glost propagator leads to the exponential fall-off of the high-energy hadron scattering amplitude at large impact parameters needed to satisfy the Froissart bound; this can solve the long-standing problem of the perturbative approach in describing high energy scattering \cite{Kovner:2001bh}. In QCD amplitudes  the coupling to the ghost can give rise to spin asymmetries \cite{Kang:2010qx} that are different from the usual perturbative approach   -- it will be interesting to study the resulting implications for spin physics at colliders.  
  \vskip0.3cm
  We thank F. Loshaj and E. Shuryak for discussions, and M. Chernodub, G. Sterman, G. Veneziano, I. Zahed and {\mbox{A. Zhitnitsky}} for useful comments. The work was supported in part by the U.S. Department of Energy under Contracts
DE-FG-88ER40388 and DE-SC0012704 (D.K.) and by the BSF grant 2012124 and the  Fondecyt (Chile) grant  1140842 (E.L.).
 \vfil
 \end{document}